\documentclass[
 reprint,
 amssymb, amsmath,
 aip,jcp
]{revtex4-1}

\usepackage{comment}
\usepackage{dcolumn}
\usepackage[colorlinks=true,linkcolor=blue]{hyperref}%
\expandafter\ifx\csname package@font\endcsname\relax\else
 \expandafter\expandafter
 \expandafter\usepackage
 \expandafter\expandafter
 \expandafter{\csname package@font\endcsname}%
\fi
\usepackage{xspace}

\usepackage{graphicx}
\usepackage{amsmath}

\usepackage[normalem]{ulem}
\usepackage[utf8]{inputenc}

\usepackage{tikz}
\usepackage{multirow}
\usetikzlibrary{positioning}

\newcommand*{\citen}{}
\DeclareRobustCommand*{\citen}[1]{%
	\begingroup
	\romannumeral-`\x 
	\setcitestyle{numbers}%
	\cite{#1}%
	\endgroup
}

\newcommand{\ket}[1]{{\ensuremath{|#1\rangle}\xspace}}

\newcommand{\basis}[0]{\mathcal{B}}

\newcommand{\psibasis}[0]{\Psi^{\basis}}

\newcommand{\br}[1]{{\mathbf{r}_{#1}}}

\newcommand{\brb}[1]{{\bf r}_{#1}}
\newcommand{\rab}[1]{|\brb{1} - \brb{2}|}

\newcommand{\ai}[1]{\hat{a}_{#1}}

\newcommand{\wbasiscoal}[1]{W_{\psibasis}(\br{},\br{})}

\newcommand{\etal}{\emph{et al.}}
\newcommand{\ku}[2]{\hat{K}[u](\br{#1},\br{#2})}
\newcommand{\kmu}[2]{\hat{K}[\mu](\br{#1},\br{#2})}
\newcommand{\lu}[3]{\hat{L}[u](\br{#1},\br{#2},\br{#3})}
\newcommand{\lmu}[3]{\hat{L}[\mu](\br{#1},\br{#2},\br{#3})}
\newcommand{\uu}[2]{u(\br{#1},\br{#2})}
\newcommand{\umu}[2]{u(r_{#1#2},\mu)}
\newcommand{\tu}{\hat{\tau}_u}
\newcommand{\tmu}{\hat{\tau}_\mu}
\newcommand{\deriv}[3]{\frac{\partial^{#3} #1}{\partial {#2}^{#3}}}

\newcommand{\wmuijkl}[4]{{w}_{ij}^{kl}}
\newcommand{\kijkl}[0]{{K}_{ij}^{kl}}
\newcommand{\lmuijmkln}[0]{{L}_{ijm}^{kln}}
\newcommand{\actset}[0]{\mathcal{A}}
\newcommand{\corset}[0]{\mathcal{C}}
\newcommand{\hmu}[0]{\tilde{H}[\mu]}
\newcommand{\hu}[0]{\tilde{H}[u]}
\newcommand{\hub}[0]{\tilde{H}[u]^\basis}
\newcommand{\hubfc}[0]{\tilde{H}[u]^\basis_{\text{FC}}}
\newcommand{\hmub}[0]{\tilde{H}[\mu]^\basis}
\newcommand{\adi}[1]{a^{\dagger}_{#1}}
\newcommand{\phiu}[0]{\ket{{\Phi_0^\basis}[u]}}
\newcommand{\phimu}[0]{\ket{{\Phi_0^\basis}[\mu]}}
\newcommand{\eob}[0]{\tilde{E}_0^\basis[u]}
\newcommand{\eomub}[0]{\tilde{E}_0^\basis[\mu]}
\newcommand{\tcfcimu}[0]{\mu\text{-TC}}
\newcommand{\tcfcihint}[0]{\text{SM-17}}

\usepackage{csquotes}

\begin{document}

\author{Werner Dobrautz}
\email{dobrautz@chalmers.se}
\affiliation{%
	Max Planck Institute for Solid State Research, Heisenbergstr. 1, 70569 Stuttgart, Germany
}%
\affiliation{
	Department of Chemistry and Chemical Engineering,
	Chalmers University of Technology, Kemig\aa rden 4, 41258 Gothenburg, Sweden
}
\author{Aron J. Cohen}
\affiliation{DeepMind, 6 Pancras Square, London N1C 4AG, UK}
\affiliation{Max Planck Institute for Solid State Research, Heisenbergstr. 1, 70569 Stuttgart, Germany}

\author{Ali Alavi}
\affiliation{%
	Max Planck Institute for Solid State Research, Heisenbergstr. 1, 70569 Stuttgart, Germany
}%
\affiliation{Yusuf Hamied Department of Chemistry, University of Cambridge, Lensfield Road, Cambridge CB2 1EW, United Kingdom}%
\author{Emmanuel Giner}%
\email{emmanuel.giner@lct.jussieu.fr}
\affiliation{Laboratoire de Chimie Théorique, Sorbonne Université and CNRS, F-75005 Paris, France}

\title{Performance of a one-parameter correlation factor for transcorrelation: the Li--Ne total energies and ionization potentials}

\newcommand{\todo}[1]{{\color{red}TODO:#1}}

\begin{abstract}
In this work we investigate the performance of a recently proposed transcorrelated (TC) approach based
on a single-parameter correlation factor \href{https://doi.org/10.1063/5.0044683}{[JCP, 154, 8, 2021]} for systems involving more than two electrons.
The benefit of such an approach relies on its simplicity as efficient numerical-analytical schemes can be set
up to compute the two- and three-body integrals occuring in the effective TC Hamiltonian.
To obtain accurate ground state energies within a given basis set, the present TC scheme is coupled to the recently proposed TC--full configuration interaction quantum Monte Carlo method \href{https://doi.org/10.1063/1.5116024}{[JCP, 151, 6, 2019]}.
We report ground state total energies on the Li--Ne series, together with their first cations, computed in increasing large basis
sets and compare to more elaborate correlation factors involving electron-electron-nucleus coordinates.
Numerical results on the Li--Ne ionization potentials show that the use of the single-parameter correlation factor brings on average only a slightly lower accuracy (1.2 mH) in a triple-zeta quality basis set with respect to a more sophisticated correlation factor.
However, already using a quadruple-zeta quality basis set yields results within chemical accuracy to complete basis set limit results when using this novel single-parameter correlation factor.

\end{abstract}

\maketitle

\section{Introduction}
At the heart of quantum chemistry lies the accurate description of the electronic structure of molecular systems,
which is a very challenging task since the corresponding mathematical problem to be solved scales exponentially with the system size.
Wave function theory (WFT) aims at solving the Schrödinger equation for a general molecular system
and provides a systematic way of improving the accuracy of the computed properties following a two-fold path: \textbf{(i)} improving the quality of the wave function in a given basis to get as close as possible from the full-configuration interaction (FCI), \textbf{(ii)} improving the quality of the one-electron basis set used to project the Schrödinger equation.
The exact properties of the system would be obtain with the FCI wave function in a complete basis set (CBS).
There exists many different flavours of wave function ansätze which approximate the FCI wave function and energy,
and they all have an unfavorable computational scaling with the  system size and most importantly, with the size of the basis set in common.
Therefore a major drawback of WFT is the slow convergence of the results with respect to the basis set size,
which mainly originates from the poor description of two-body density matrix near the electron-electron coalescence point (\textit{i.e.} $r_{12}\approx 0$).

A central idea shared by the theories aiming to improve the convergence of WFT with respect to the basis set is related to the so-called
electron-electron cusp-condition derived by Kato\cite{Kat-CPAM-57}:
one multiply the cusp-less wave function developed in an incomplete basis set by a correlation factor explicitly depending on the $r_{12}$ coordinate which restores the cusp-condition.
Beside the cusp, the most important role of the correlation factor is to lower the probability of finding two electrons near one another, which is often referred as digging the short-range part of the Coulomb hole.

There are mainly three classes of theories dealing with a correlation factor:
\textbf{(i)} F12 theory\cite{Ten-TCA-12,TenNog-WIREs-12,HatKloKohTew-CR-12, KonBisVal-CR-12, GruHirOhnTen-JCP-17, MaWer-WIREs-18} where one projects out the effect of the correlation factor from the incomplete basis set used to compute the cusp-less wave function,
\textbf{(ii)} variational Monte Carlo (VMC) methods\cite{TouAssUmr-book-vmc} where the full effect of the correlation factor is retained in the wave function and all parameters are variationally optimized and
\textbf{(iii)}  the transcorrelated theory (TC) \cite{Hirschfelder-JCP-63,BoyHan-PRSLA-69,BoyHan-2-PRSLA-69} where the effect of the full correlation factor is incorporated through a non hermitian effective Hamiltonian.
All these three theories have been shown to strongly reduce the basis set convergence problem of WFT.

The main advantage of the TC theory is that it combines favourable aspects of both VMC and F12: \textbf{(i)} since only up to effective three-electron terms are needed (compared to the N$-$body terms of VMC), usual post-Hartree Fock methods can be designed to solve the TC Hamiltonian, \textbf{(ii)} no more than $\mathbb{R}^6$ integrals are needed, and \textbf{(iii)} compact wave function can be obtained because the full correlation factor is taken into account.
Despite these attractive features, the two main drawbacks of the TC theory are that \textbf{(i)}  the non-hermitian nature of the TC operator which induces the loss of variationality, and \textbf{(ii)}  that the three-body terms generate an $M^6$ tensor -- $M$ is the number of basis set functions -- which becomes rapidly prohibitive to store during calculations.
Nevertheless, because it originates from a similarity transformation, the exact eigenvalues are obtained when reaching the CBS limit, which suggests that the loss of variational property in TC is a signature of a too constrained form of the wave function.
Regarding the functional form of the cusp-less wave function and of the correlation factors, 
the seminal work of Boys and Handy\cite{BoyHan-PRSLA-69,BoyHan-2-PRSLA-69} proposed to optimize both the orbitals of a single Slater determinant and a sophisticated correlation factor. 
Then, Ten-No\cite{TenNo-CPL-00-a} proposed to significantly change of paradigm since he used the combination of a rather simple universal correlation factor whose shape was optimized for the range of valence electrons, and a rather sophisticated ansatz for the wave function  (M{\o}ller-Plesset at second order in Refs. \onlinecite{TenNo-CPL-00-a,HinTanTen-JCP-01} and  linearised coupled-cluster in Ref. \cite{HinTanTen-CPL-02}).
The works of Ten-No have shown a faster convergence of the TC theory (such as TC-MP2) towards the exact energies with respect to the basis set with 
respect to their parent usual WFT theory (such as regular MP2). 
Nevertheless, it should be mentioned that because the correlation factor was optimized for valence electrons,
the use of basis sets explicitly optimized for core electrons (\textit{e.g.} the cc-pCVXZ family) is mandatory
in order to maintain a sensible value for the energy in all-electron calculations.

More recently, Cohen \textit{et. al}\cite{CohLuoGutDobTewAla-JCP-19} applied the TC methodology with an elaborate correlation factor,
and proposed to use the full configuration interaction Monte Carlo (FCIQMC) method to obtain the exact ground state energy and the corresponding right eigenvector of the TC Hamiltonian in given basis set.
In their work\cite{CohLuoGutDobTewAla-JCP-19}, the authors used the Jastrow factors of Moskowitz \textit{et. al.}\cite{SchMos-JCP-90} 
optimized in the context of VMC for the He-Ne neutral series, and which explicitly take into account electron-electron-nucleus (e-e-n) correlation effects. 
This work has showed the beneficial impact of the e-e-n terms in yielding highly accurate total energies and ionisation potentials using the TC-FCIQMC method with modest basis sets. The TC-FCIQMC method has also been applied to the binding curve of the Be$_2$ system, yielding spectroscopic accuracy across the entire binding curve using only triple-zeta basis sets, demonstrating how the TC-FCIQMC method can be used in ab initio problems with a delicate balance between static and dynamical correlation \cite{GutCohLuoAla-JCP-21}.  In other work from the Alavi group, the application of the TC-FCIQMC method to the 2D Hubbard model\cite{DobLuoAla-PRB-19} showed how transcorrelation can be beneficial in the treatment of {\em strongly correlated} systems, by compressing the right eigenvector of the ground state so that it becomes largely dominated by the Hartree-Fock determinant, in a regime where the ground-state eigenvector of the non-transcorrelated Hubbard Hamiltonian is strongly multi-configurational. Similar application of the Gutzwiller Ansatz\cite{Gutzwiller-PRL-63,BriRic-PRB-70} was recently reported by Reiher \textit{et. al.}\cite{BaiRei-JCP-20} using density matrix renormalisation group
and various methods based on the TC approach have been used to reduce the resource requirements for accurate 
electronic structure calculations on state-of-the-art quantum computing hardware\cite{ValTak-PCCP-20, McArdle2020, Schleich2021, Kumar2022, Sokolov2022}. 

Recently, one of the present authors introduced a single-parameter correlation factor\cite{Gin-JCP-21} inspired by range-separated density functional theory (RS--DFT).
The main idea developed in this work was to find a mapping between the leading order terms in $1/r_{12}$ of the effective scalar potential obtained in the TC equations and the non divergent long-range interaction $\text{erf}(\mu r_{12})/r_{12}$ used in RS--DFT. The correlation factor obtained with such a procedure has an explicit analytical form which depends on a single parameter $\mu$: the lower the $\mu$, the deeper is the correlation hole dug by the correlation factor, and in the $\mu \rightarrow \infty$ limit the effect of the correlation factor vanishes.
Preliminary tests on atomic and molecular two-electron systems have shown that this TC framework also improves the convergence of the energy, and that a good value of the parameter $\mu$ could be systematically obtained with nothing more than the knowledge of the Hartree-Fock (HF) density.
The advantage of this simple correlation factor is that the corresponding TC Hamiltonian has a rather simple analytical form for which the two- and three-body integrals can be very efficiently obtained using a mixed numerical and analytical scheme.

The aim of the present work is to study how this relatively simple correlation factor performs for systems with more than two electrons.
In order to be able to eliminate any source of errors within a basis set, we use the FCIQMC approach to obtain the exact right eigenvector in a given basis set.
We are then able to compare with the results obtained with the more sophisticated correlation factor used in the recent work of Cohen \textit{et. al.}.\cite{CohLuoGutDobTewAla-JCP-19}

The remainder of this article is organized as follows:
In Section~\ref{sec:theo}, we recap the main equations of the TC theory together with the explicit form of the TC Hamiltonian obtained
in Ref.~\onlinecite{Gin-JCP-21}.
Then, in Section~\ref{sec:results_BNe} we investigate the sensitivity of the present approach with the quality of description of core electrons and in Section~\ref{sec:results_5idx} we investigate a possible approximation to the numerous $N^6$ tensor of the three-body integrals inherent to the TC approach within the Ansatz considered here.
In Section~\ref{sec:results_full} we report the results for total energies of neutral and first cation species on the Li--Ne series as a function of increasing basis set size.
We also compare the quality of the ionization potentials (IPs) obtained with the present approach with the existing literature.
Eventually we conclude in Section~\ref{sec:conclusion}.

\section{Theory}
\label{sec:theo}
\subsection{General equations and concepts of TC theory}
The general form of the transcorrelated Hamiltonian for a symmetric correlation factor $\uu{1}{2}$ is given by
\begin{equation}
 \label{ht_def_g}
 \begin{aligned}
  \hu  &\equiv e^{-\tu} \hat{H} e^{\tu} \\
                & = H + \big[ H,\tu \big] + \frac{1}{2}\bigg[ \big[H,\tu\big],\tu\bigg],
 \end{aligned}
\end{equation}
where $\tu = \sum_{i<j}u(\br{i},\br{j})$ and $\hat{H} = -\sum_i \frac{1}{2} \nabla^2_i + v(\br{}_i) + \sum_{i<j}   \frac{1}{r_{ij}}$.
Eq. \eqref{ht_def_g} leads to the following transcorrelated Hamiltonian
\begin{equation}
 \begin{aligned}
 \label{ht_def_g2}
 \hu& = H - \sum_{i<j} \ku{i}{j} - \sum_{i<j<k} \lu{i}{j}{k}
 \end{aligned}
\end{equation}
where the effective two- and three-body operators $\ku{1}{2}$ and $\lu{1}{2}{3}$ are defined as
\begin{equation}
 \begin{aligned}
  \ku{1}{2} = \frac{1}{2} \bigg( &\Delta_1 \uu{1}{2} + \Delta_2 \uu{1}{2} \\
                                         + &\big(\nabla_1 \uu{1}{2} \big) ^2 + \big(\nabla_2 \uu{1}{2}      \big) ^2 \bigg) \\
                                         + &\nabla_1 \uu{1}{2} \cdot \nabla_1 + \nabla_2 \uu{1}{2}\cdot     \nabla_2
 \end{aligned}
\end{equation}
and
\begin{equation}
 \begin{aligned}
  \lu{1}{2}{3} = &   \nabla_1 \uu{1}{2} \cdot \nabla_1 \uu{1}{3} \\
                                          + & \nabla_2 \uu{2}{1} \cdot \nabla_2 \uu{2}{3}   \\
                                          + & \nabla_3 \uu{3}{1} \cdot \nabla_3 \uu{3}{2}   .
 \end{aligned}
\end{equation}
In practice, the TC Hamiltonian is projected into a basis set $\basis$
\begin{equation}
 \label{eq:def_hub}
 \begin{aligned}
 &\hub   = P^{\basis} \hu P^\basis,
 \end{aligned}
\end{equation}
where $P^{\basis}$ is the projector onto a given basis set $\basis$.
Using real-valued orthonormal spatial molecular orbitals (MOs) $\{\phi_i(\br{})\}$, $\hub$ can be written in a second-quantized form  as
\begin{equation}
 \label{eq:def_hub_sec_q}
 \begin{aligned}
 &\hub  = \sum_{i,j \in \basis} \,\, \sum_{\sigma = \uparrow,\downarrow} h_{ij} \adi{j,\sigma}\ai{i,\sigma}\\ 
 & + \frac{1}{2}\sum_{i,j,k,l \in \basis}  \,\, \sum_{\sigma,\lambda = \uparrow,\downarrow}
 \big( V_{ij}^{kl} - \kijkl\big) \adi{k,\sigma} \adi{l,\lambda} \ai{j,\lambda} \ai{i,\sigma} \\
       & - \frac{1}{6} \sum_{i,j,m,k,l,n \in \basis} \,\, \sum_{\sigma,\lambda,\kappa = \uparrow,\downarrow}
\lmuijmkln \adi{k,\sigma} \adi{l,\lambda} \adi{n,\kappa} \ai{m,\kappa} \ai{j,\lambda} \ai{i,\sigma}
 \end{aligned}
\end{equation}
where $h_{ij}$ are the usual one-electron integrals, $V_{ij}^{kl}$ are the usual two-electron integrals,
$\kijkl$ are the two-electron integrals corresponding to the effective two-body operator $\ku{1}{2}$ operator
\begin{equation}
 \kijkl = \int \text{d} \br{1} \text{d} \br{2} \phi_k(\br{1}) \phi_l(\br{2}) \ku{1}{2} \phi_i(\br{1}) \phi_j(\br{2}),
\end{equation}
and $\lmuijmkln$ are the three-electron integrals corresponding to the effective three-body operator $\lu{1}{2}{3}$
\begin{equation}
 \begin{aligned}
  \lmuijmkln  = \int \text{d} \br{1} \text{d} \br{2} \text{d} & \br{3} \phi_k(\br{1}) \phi_l(\br{2}) \phi_n(\br{3}) \\ & \lu{1}{2}{3} \phi_i(\br{1}) \phi_j(\br{2}) \phi_m(\br{3}).
 \end{aligned}
\end{equation}
The ground state eigenvalue and the associated right eigenvector fulfill the eigenvalue equation
\begin{equation}
 \hub \phiu = \eob \phiu,
\end{equation}
and because of the properties of the similarity transformation the exact ground state energy $E_0$ is recovered in the CBS limit
\begin{equation}
 \lim_{\basis \rightarrow \text{CBS}} \eob = E_0,
\end{equation}
for all correlation factors $u(\br{i},\br{j})$ chosen to obtain $\hub$.
If $u(\br{i},\br{j})$  is properly chosen one expects a fast convergence of $\eob$ towards $E_0$.
Nevertheless, because of the loss the of variational principle of $\eob$ due to the non hermitian character of $\hub$, this convergence is not guaranteed to be monotonic as in the usual WFT calculations, and $\eob$ can be bellow the exact ground state energy.

\subsection{One-parameter TC Hamiltonian: $\hmu$}
Recently, one of the present authors\cite{Gin-JCP-21} have introduced a one-parameter correlation factor $\umu{1}{2}$
based on a mapping between the $r_{12} \approx 0$ limit of the TC Hamiltonian and the range separated DFT effective Hamiltonian.
The explicit form of $\umu{1}{2}$ derived in Ref. \onlinecite{Gin-JCP-21} reads  as
\begin{equation}
 \label{eq:def_j}
 \umu{1}{2} = \frac{1}{2}r_{12}\bigg( 1 - \text{erf}(\mu r_{12})  \bigg) - \frac{1}{2\sqrt{\pi}\mu}e^{-(r_{12}\mu)^2}.
\end{equation}
Because of the simple analytical expression of $\umu{1}{2}$, the corresponding TC Hamiltonian  $\hmu$ defined as
\begin{equation}
 \label{eq:def_h_mu}
 \begin{aligned}
 \hmu &\equiv e^{-\tmu} \hat{H} e^{\tmu} \\
      & = H - \sum_{i<j} \kmu{i}{j} - \sum_{i<j<k} \lmu{i}{j}{k},
 \end{aligned}
\end{equation}
with $\tmu = \sum_{i<j}\umu{i}{j}$, has a relatively simple analytical form
with the effective two- and three-body operators
\begin{equation}
 \label{eq:k_final}
  \begin{aligned}
   & \kmu{i}{j}=  \frac{1 - \text{erf}(\mu r_{12})}{r_{12}} - \frac{\mu}{\sqrt{\pi}} e^{-\big(\mu       r_{12} \big)^2} \\
                &+ \frac{\bigg(1 -     \text{erf}(\mu r_{12}) \bigg)^2}{4} - \bigg( \text{erf}(\mu r_{12}) - 1\bigg) \deriv{}{r_{12}}{}
  \end{aligned}
\end{equation}
and
\begin{equation}
 \label{eq:l_final}
 \begin{aligned}
 \lmu{i}{j}{k} = & \frac{1 - \text{erf}(\mu r_{12})}{2 r_{12}} \br{12} \cdot \frac{1 -            \text{erf}(\mu r_{13})}{2 r_{13}} \br{13} \\
                                      + & \frac{1 - \text{erf}(\mu r_{12})}{2 r_{12}} \br{21} \cdot \frac{1 -            \text{erf}(\mu r_{23})}{2 r_{23}} \br{23} \\
                                      + & \frac{1 - \text{erf}(\mu r_{13})}{2 r_{13}} \br{31} \cdot \frac{1 -            \text{erf}(\mu r_{32})}{2 r_{32}} \br{32},
 \end{aligned}
\end{equation}
respectively.
The correlation factor $\umu{1}{2}$ exactly restores the cusp conditions and the effective Hamiltonian $\hmu$ obtained with the scalar two- and three-body effective interaction in Eqs.~\eqref{eq:k_final} and~\eqref{eq:l_final} is non divergent, yielding \enquote{cusp-less} eigenvectors as illustrated in Ref~\onlinecite{Gin-JCP-21}.
As apparent from the definitions of Eq~\eqref{eq:k_final} and Eq.~\eqref{eq:l_final}, the global shape of $\hmu$ depends on a unique parameter $\mu$, which can be seen either as the inverse of the typical range of the correlation effects, or the typical value of the effective interaction at $r_{12}=0$.
In the $\mu \rightarrow +\infty$ limit one obtains the usual Hamiltonian, while in $\mu \rightarrow 0 $ limit one obtains an attractive non hermitian Hamiltonian.

Similarly to Eq.~\eqref{eq:def_hub}, we define the projection onto a basis set $\basis$ of the TC Hamiltonian $\hmu$
\begin{equation}
 \hmub \equiv P^\basis \hmu P^\basis,
\end{equation}
whose ground state eigenvalue and associated right-eigenvector satisfy
\begin{equation}
 \hmub \phimu = \eomub \phimu.
\end{equation}

Because of its relatively simple form, the correlation factor $\umu{1}{2}$ has the advantage that it leads
to effective operators $\kmu{1}{2}$ and $\lmu{1}{2}{3}$ with a simple-enough analytical form for which integrals
can be computed efficiently using a mixed numerical-analytical scheme  (see Ref \onlinecite{Gin-JCP-21} for explicit formulas).
This is in contrast to the ${\mathbb R}^6$ numerical integrals needed when using more sophisticated correlation factors.

\section{Results}
\label{sec:results}

\subsection{Computational details}
To obtain the ground state eigenvalue $\eob$ of a given TC Hamiltonian $\hub$, we use the recently developed similarity transformed full-configuration interaction quantum Monte Carlo (ST-FCIQMC) technique\cite{CohLuoGutDobTewAla-JCP-19, GutCohLuoAla-JCP-21} which extends
the original stochastic projection technique of FCIQMC\cite{BooThoAla-JCP-09, BooAla-JCP-10, BooCleThoAla-JCP-11, GhaLozAla-JCP-19, VitAlaKat-JCTC-20, Guther2020, Dobrautz2019, Dobrautz2021}
to a non-hermitian and three-body Hamiltonian. 
The FCIQMC parameters were $10^6$ walkers, an initiator threshold of $n_{init} = 3$ and a semi-stochastic space of $N_D = 1000$.
Provided a $\hub$ and a given basis set $\basis$, the necessary one-, two- and three-body integrals are computed using restricted Hartree-Fock (RHF) MOs.
When the correlation factor is $\umu{1}{2}$, we label the results by $\tcfcimu$, whereas when using the correlation factor of Moskowitz \textit{et. al.}\cite{SchMos-JCP-90} we label the results by $\tcfcihint$.
Regarding the integrals involved in $\hmub$, the scalar two-body part is computed analytically and the non hermitian together with the three-body parts are computed using a mixed analytical-numerical scheme where the Becke's numerical grid\cite{Bec-JCP-88b} contains 30 radial points and a Lebedev angular grid of 50 grid points. Numerical tests have shown that these relatively small number of grid points ensures a sub $\mu\text{Ha}$ convergence of the total energies.

Regarding the value of $\mu$ chosen here, thorough this article we use the so-called RSC+LDA system-dependent value defined in Eq. (57) of Ref. \onlinecite{Gin-JCP-21} as such a strategy was found to be the most accurate on the study of two-electron systems in the latter work. 
Estimates of the FCI in a given basis set $\basis$ and within a sub mH precision were obtained with the configuration interaction perturbatively selected iteratively\cite{malrieu_cipsi} (CIPSI) as implemented in the Quantum Package\cite{QP2}. 
The estimated CBS all-electron results for atoms and cations are taken from Ref. \onlinecite{ChaGwaDavParFro-PRA-93}.
Except for the ST-FCI-QMC, all calculations where performed using the Quantum Package\cite{QP2}.

\subsection{Preliminary investigation on B, B$^+$, Ne and Ne$^+$ }
\label{sec:results_BNe}
Before performing the study on the whole Li--Ne series together with their first cations, we perform a detailed study on the neutral and first cations of the boron and neon atoms.
The main questions we address are \textbf{(i)} how to treat core electrons in all-electron calculations using $\hmu$ and \textbf{(ii)} to investigate a possible reduction of the computational cost involved in the three-body operator while maintaining the accuracy.
\subsubsection{Treatment of core electrons in all-electron calculations}
We begin our preliminary investigation by studying the treatment of core electrons with $\hmu$ in the case of the boron neutral atom.
We performed all electron $\tcfcimu$ calculations with the cc-pVXZ and cc-pCVXZ basis sets (X=D,T) to study the impact of functions suited for core-valence correlation, and we report the results in Table \ref{tab:no-core-mu}.
From Table \ref{tab:no-core-mu} we can observe that the all electron $\tcfcimu$ calculations without core-valence functions significantly underestimate the exact ground state energy of the boron atom by 36 mH and 26 mH in the cc-pVDZ and cc-pVTZ basis sets, respectively.
On the other hand such effect is strongly reduced when using core-valence functions as the underestimation of the ground state energy is of 2.2 mH and 1.8 mH with the cc-pCVDZ and cc-pCVTZ basis sets, respectively. 
Regarding now the effect on the IP, it can be noticed that while the energy difference computed using a cc-pCVDZ is already 
within a sub mH precision with respect to the CBS value, the results obtained without the core-valence functions are far from such an accuracy as the error is of 17 and 15 mH using the cc-pVDZ and cc-pVTZ, respectively.  
Therefore, as already shown by Ten-No \textit{et. al.}\cite{HinTanTen-JCP-01},
core-valence correlation functions are mandatory when performing all electron calculations in the context of TC methods,
unless when the correlation factor includes explicit electron-electron-nucleus correlation
as for instance in the work of Cohen \textit{et. al.}\cite{CohLuoGutDobTewAla-JCP-19}.
\begin{table}
 \label{table_b}
\caption{\label{tab:no-core-mu}Boron and boron cation total energies results (in a.u.) and ionization potentials (IP) from all-electron calculations for $\mu_{\text{RSC+LDA}} \approx 1.02$ for Boron and $\mu_{\text{RSC+LDA}} \approx 1.15$ for the boron cation with and without core-valence basis functions. }
\begin{tabular}{ccccc}
\toprule
Atom  & & DZ & TZ & CBS\protect\cite{ChaGwaDavParFro-PRA-93, DavHagChaMeiFro-PRA-91} \\
 \hline
\multirow{2}{*}{B} & with core-valence & -24.65613 & -24.65568 & \multirow{2}{*}{-24.65391} \\
 				   & w/o core-valence  & -24.69075 & -24.68063 & \\
 \hline
 \multirow{2}{*}{B$^+$} & with core-valence & -24.35147 & -24.34960 & \multirow{2}{*}{-24.34889} \\
 						& w/o core-valence  & -24.37284 & -24.36495 & \\
 \hline
\multirow{2}{*}{IP} & with core-valence & 0.30466 & 0.30608  & \multirow{2}{*}{0.30502} \\
 				   & w/o core-valence  & 0.31791 & 0.31568 & \\
 \botrule
\end{tabular}
\end{table}

\subsubsection{The \enquote{5-idx} approximation on the three-body term}
\label{sec:results_5idx}
Another important computational aspect of the TC method are the numerous $N^6$ integrals to be computed for the three-body effective operator $\lu{1}{2}{3}$.
The problems regarding these terms are two-fold: \textbf{(i)} the computation of the intermediate quantities which can be quite demanding and \textbf{(ii)} the computation and storage of all the $N^6$ integrals.
In the context of $\hmub$, point \textbf{(i)} is not really a problem since all intermediate quantities are computed analytically and not numerically in contrast to more complex correlation factors\cite{CohLuoGutDobTewAla-JCP-19, GutCohLuoAla-JCP-21} (see Appendix of Ref.~\onlinecite{Gin-JCP-21}).
Therefore, the main computational bottleneck is the computation and storage of the $N^6$ integrals.
Nevertheless, one can notice that the most numerous terms in the $\lmuijmkln$ tensor are those corresponding to 6 different indices, which corresponds to pure triple excitations operators. 
We propose here the \enquote{5-idx} approximation of the three-body term which consists in neglecting all integrals $\lmuijmkln$ with six different indices, which reduces to $N^5$ the number of integrals to compute and store for the treatment of $\lu{1}{2}{3}$.
We performed numerical calculation with the full treatment of the $\lu{1}{2}{3}$ operator and the \enquote{5-idx} approximation using the cc-pCVXZ (X=D,T,Q) for the neon atom and its first cation, and report the results in Table~\ref{tab:5idx}.
From Table~\ref{tab:5idx} we can observe that the results are almost insensitive to the 5-idx approximation as the differences between the energies are of about 10$^{-5}$H for both the neon and the first cation.
This result therefore indicates that the 5-idx approximation drastically reduces both the memory and CPU bottleneck of the TC calculations
while leaving the numerical results unchanged to a sub $m\text{Ha}$ precision.
This is nevertheless still more expensive than the normal-ordered approaches proposed by some of the present authors in Ref. \onlinecite{SchCohAla-JCP-21}, but has nevertheless the advantage not to depend on the one- and two-body density of some reference wave function.
\begin{table}
	\small
	\renewcommand{\arraystretch}{1.2}
\caption{\label{tab:5idx}Effect of neglecting the full 3-body terms on total energies (reported in a.u.) for all electron calculations in $\tcfcimu$ in a cc-pCVXZ basis sets (X=D,T,Q) for Neon and Ne$^+$ and the corresponding ionization potential (reported in m a.u.). 
}
\begin{tabular}{ccccc}
\toprule
Method & Basis & Ne [H] & Ne$^+$ [H] & IP [mH] \\
\hline
Full & cc-pCVDZ & -128.96435(1)\phantom{0} 			& -128.16345(1)\phantom{0} 	& 800.90(2)\phantom{0} \\
5idx & cc-pCVDZ & -128.96437(2)\phantom{0} 			& -128.16344(2)\phantom{0} 	& 800.93(4)\phantom{0} \\
Full & cc-pCVTZ & -128.93230(1)\phantom{0} 			& -128.14083(8)\phantom{0} 	& 791.47(9)\phantom{0} \\
5idx & cc-pCVTZ & -128.93201(2)\phantom{0} 			& -128.14064(6)\phantom{0} 	& 791.38(8)\phantom{0} \\
Full & cc-pCVQZ & -128.93569(3)\phantom{0} 			& -128.14295(8)\phantom{0}  & 792.7(1)\phantom{00}\\
5idx & cc-pCVQZ & -128.93545(2)\phantom{0}			& -128.14264(1)\phantom{0}  & 792.81(3)\phantom{0}\\
\hline
\botrule
\end{tabular}
\end{table}

\subsection{All electrons calculations the Li-Ne species and first cations}

\begin{table*}
	\caption{\label{tab:full-core-a}Ground state all electron calculations for the Li--Ne species, together with their first cations and the corresponding ionization potential (IP) computed in the cc-pCVXZ (X=D,T,Q) family of basis set. SM-17 results stand for TC-FCIQMC with the correlation factor of Moskowitz \textit{et. al.}\cite{SchMos-JCP-90} to obtain the TC Hamiltonian. 
	Estimated non relativistic CBS results are obtained from Ref.~\onlinecite{ChaGwaDavParFro-PRA-93}. All results are reported in atomic units.}
	\renewcommand{\arraystretch}{1.1}
	\begin{tabular}{lccccc|cccccc}
		\toprule
		Atom    &Method       &\multicolumn{1}{c}{CVDZ}&\multicolumn{1}{c}{CVTZ}&  \multicolumn{1}{c}{CVQZ}& \multicolumn{1}{c}{Est. CBS$^a$} & Atom    &Method       &\multicolumn{1}{c}{CVDZ}&\multicolumn{1}{c}{CVTZ}&  \multicolumn{1}{c}{CVQZ}& \multicolumn{1}{c}{Est. CBS$^a$} \\
\hline
		&$\tcfcimu$ & -7.47909  & -7.47857  & -7.47832 & 			&		&$\tcfcimu$	& -54.59896 & -54.59382 & -54.59019 & 			\\
Li 		& SM-17	    & -7.47748  & -7.47824  &\textendash& -7.47806	& N		& SM-17		& -54.56695 & -54.58658 &\textendash&  -54.58920			\\
		& CIPSI     & -7.46602  & -7.47424  & -7.47636 &            & 		& CIPSI     & -54.51765 & -54.56793 & -54.58197 & \\[3pt]
		&$\tcfcimu$ & -7.27988  & -7.28003  & -7.27999 & 			&		&$\tcfcimu$ & -54.05787 & -54.05721 & -54.05478 & 			\\
Li$^+$  & SM-17     & -7.27951  & -7.28016 &\textendash& -7.27991   & N$^+$ & SM-17     & -54.03875 & -54.05159 &\textendash&   -54.05460                \\
		& CIPSI     & -7.26919  & -7.27655  & -7.27833 &            &		& CIPSI     & -53.99612 & -54.03672 & -54.04850 &  \\[3pt]
		&$\tcfcimu$ &  0.19921  &  0.19853  &  0.19833 & 			&		&$\tcfcimu$ &   0.54109 &   0.53661 &   0.53545 & 			\\
IP(Li)	& SM-17		&  0.19797  &  0.19808 &\textendash& 	0.19815	& IP(N)	& SM-17		&   0.52820 &   0.53499 &\textendash& 0.53460 \\
		& CIPSI 	&  0.19683  &  0.19769  &  0.19803 &   			& 		& CIPSI 	&   0.52153 &   0.53121 &   0.53347 &    \\
\hline
		&$\tcfcimu$ & -14.67037 & -14.66855 & -14.66777 & 			&		&$\tcfcimu$ & -75.07412 & -75.06774 & -75.06729 & 			\\
Be 		& SM-17		& -14.66969 & -14.66863	&\textendash& -14.66736	& O		& SM-17 	& -75.02676 & -75.06082 &\textendash& -75.06730			\\
		& CIPSI	    & -14.65182 & -14.66236 & -14.66556 &           &		& CIPSI	    & -74.95051 & -75.03122 & -75.05447 &  \\[3pt]
		&$\tcfcimu$ & -14.32565 & -14.32538 & -14.32504 & 			&		&$\tcfcimu$ & -74.57748 & -74.57055 & -74.56829 & 			\\
Be$^+$  & SM-17     & -14.32570 & -14.3254 &\textendash & -14.32476 & O$^+$ & SM-17     & -74.54189 & -74.56201 &\textendash& -74.56680  \\
		& CIPSI	    & -14.31102 & -14.32048 & -14.32317 &           &		& CIPSI     & -74.47796 & -74.54098 & -74.55815 &	\\[3pt]
		&$\tcfcimu$ &   0.34472 &   0.34317 &   0.34272 & 			&		&$\tcfcimu$ &   0.49664 &   0.49719 &   0.49900 & 			\\
IP(Be) 	& SM-17		&   0.34399 &   0.34323 &\textendash& 0.34258	& IP(O)	& SM-17		&   0.48487 &   0.49881 &\textendash&  0.50050 \\
		& CIPSI	    &   0.34080 &   0.34188 &   0.34239 &   	    &		& CIPSI     &   0.47255 &   0.49024 &   0.49632 &    \\
\hline
		&$\tcfcimu$ & -24.65612 & -24.65568 & -24.65467 &  			&		&$\tcfcimu$ & -99.74701 & -99.73164 & -99.73326 & 			\\
B  	 	& SM-17		& -24.65169 & -24.65459 &\textendash& -24.65391	& F		& SM-17		& -99.67001 & -99.72284 &\textendash& -99.73390\\
		& CIPSI	    & -24.62603 & -24.64485 & -24.65083 & 	        &		& CIPSI     & -99.56965 & -99.68185 & -99.71509 &  \\[3pt]
		&$\tcfcimu$ & -24.35147 & -24.34960 & -24.34953 &  			&		&$\tcfcimu$ & -99.10398 & -99.09299 & -99.09358 & 			\\
B$^+$   & SM-17     & -24.35028 & -24.34916 &\textendash& -24.34892 & F$^+$ & SM-17     & -99.04116 & -99.08228 &\textendash&   -99.09280                 \\
		& CIPSI     & -24.32985 & -24.34213 & -24.34664 &           &		& CIPSI     & -98.95526 & -99.05157 & -99.07850 &	\\[3pt]
		&$\tcfcimu$ &   0.30466 &   0.30608 &   0.30513 &    		&		&$\tcfcimu$ &   0.64303 &   0.63865 &   0.63968 & 			\\
IP(B)	& SM-17		&   0.30141 &   0.30542 &\textendash&   0.30499 & IP(F)	& SM-17		&   0.62885 &   0.64056 &\textendash&  0.64110 \\
		& CIPSI     &   0.29618 &   0.30272 &   0.30419 &           &		& CIPSI	    &   0.61439 &   0.63028 &   0.63659 &   \\
		\hline                                                                                                                         
		&$\tcfcimu$ & -37.84888 & -37.84793 & -37.84617 & 			&		&$\tcfcimu$ &-128.96435 &-128.93221 &-128.93569 & 			\\ 
C		& SM-17 	& -37.83537 & -37.84462 &\textendash& -37.84500	& Ne  	& SM-17		&-128.84774 &-128.91945 &\textendash& -128.93760  \\                 
		& CIPSI		& -37.79798 & -37.83003 & -37.83962 &           & 		& CIPSI	    &-128.72254 &-128.86823 &-128.91235 &       \\[3pt]                            
		&$\tcfcimu$ & -37.43162 & -37.43214 & -37.43170 & 			&		&$\tcfcimu$ &-128.16345 &-128.14082 &-128.14298 &			\\
C$^+$   & SM-17     & -37.42606 & -37.43018 &\textendash& -37.43103 & Ne$^+$& SM-17		&-128.06691 &-128.12553 &\textendash & -128.14310 \\                                        		
		& CIPSI	    & -37.39487 & -37.41932 & -37.42711 &			&		& CIPSI     &-127.95437 &-128.08498 &-128.12259 &   \\[3pt]
		&$\tcfcimu$ &   0.41727 &   0.41579 &   0.41447 & 			&		&$\tcfcimu$ &   0.80090 &   0.79139 &   0.79271 &  			\\ 
IP(C)	& SM-17		&   0.40931 &   0.41444 &\textendash&   0.41397 & IP(Ne)& SM-17		&   0.78083 &   0.79392 &\textendash&   0.79450  \\
		& CIPSI	    &   0.40345 &   0.41123 &   0.41303 &  			&		& CIPSI     &   0.76817 &   0.78325 &   0.78976 &                 \\
		\botrule
	\end{tabular}
\end{table*}

\begin{table}
\caption{\label{tab:mae}Mean absolute errors (MAE) in mH for the ionization potentials at the $\tcfcimu$ and CIPSI levels of theory for the Li--Ne series in the cc-pCVXZ basis sets.
The results labelled by SM-17 are the TC-FCIQMC results using of Ref.~\onlinecite{CohLuoGutDobTewAla-JCP-19} using a flexible correlation factor. }
  \begin{tabular}{lccc}
  	\toprule
              & CVDZ & CVTZ & CVQZ  \\
  \hline
  CIPSI       & 14.65& 5.24 & 2.08  \\
  $\tcfcimu$  & 3.19 & 1.85 & 0.81  \\
  SM-17       & 7.22 & 0.60 &  -    \\
  \botrule
 \end{tabular}
\end{table}

   \begin{figure*}
   \centering
\includegraphics[width=0.9\textwidth]{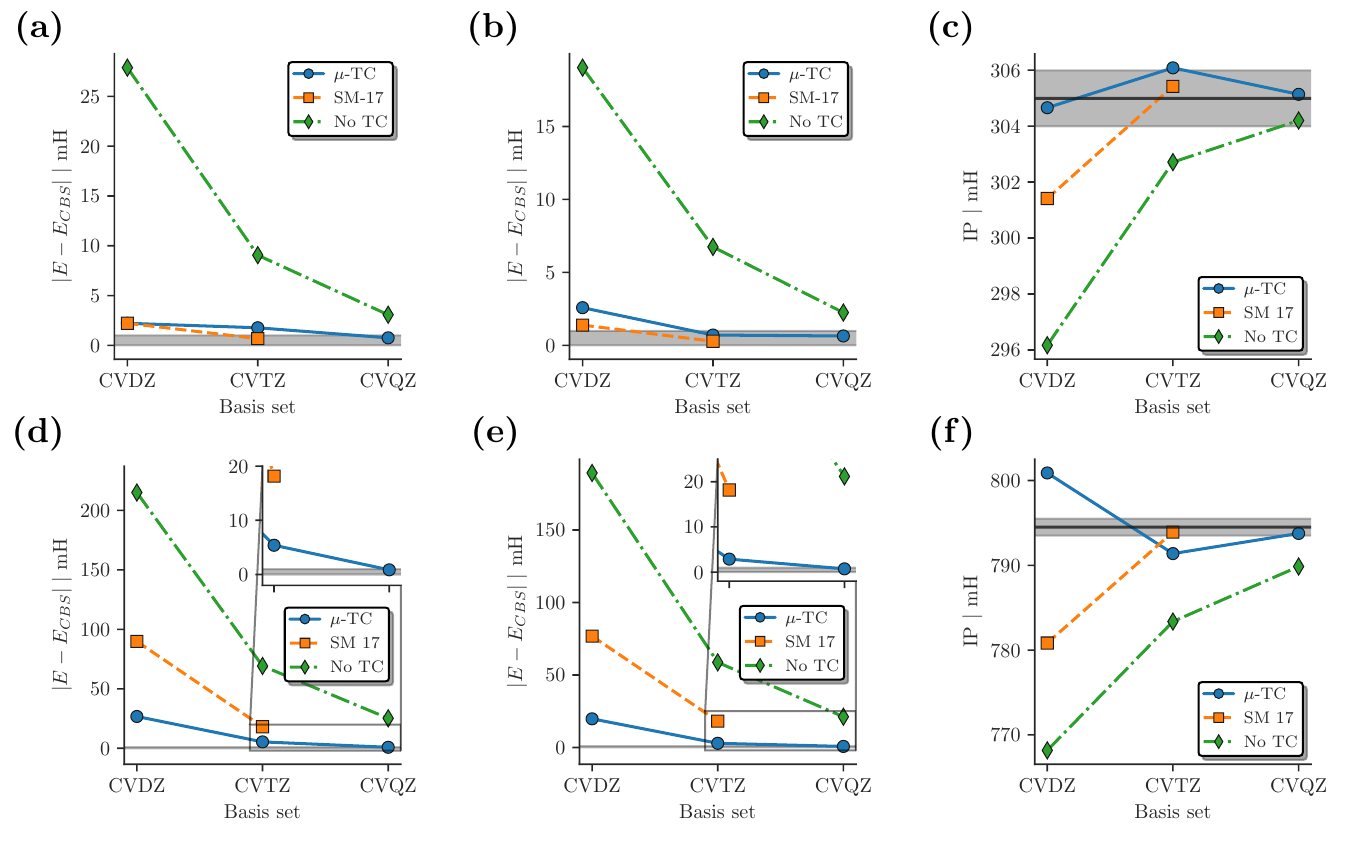}
   \caption{\label{fig:boron-neon}Boron (a), B$^+$ (b), Ne (d) and Ne$^+$ (e) absolute energy difference with respect to the estimated exact total energies\protect\cite{ChaGwaDavParFro-PRA-93} and ionization potential for boron (c) and neon(f) of the range separated TC ($\mu$-TC), and Cohen~\etal\cite{CohLuoGutDobTewAla-JCP-19} (SM-17) and non-transcorrelated results (no TC) for fully correlated calculations with cc-pCVXz core-valence basis sets.
          The gray areas indicates sub mH chemical accuracy.}
   \end{figure*}

\label{sec:results_full}
We report in Tables~\ref{tab:full-core-a} the performance of all-electron ground state calculations on the Li--Ne series in the cc-pCVXZ basis sets (X=D,T,Q), together with their first cations at the $\tcfcimu$ and CIPSI levels of theory.
We also report in Tables~\ref{tab:full-core-a} the estimated non relativistic CBS results of Ref.~\onlinecite{ChaGwaDavParFro-PRA-93}, together with the so-called $\tcfcihint$ results which are
the TC-FCIQMC calculations in the cc-pCVXZ basis set family (X=D,T) using the same methodology of Ref. ~\onlinecite{CohLuoGutDobTewAla-JCP-19}
where the correlation factor of Moskowitz \textit{et. al.}\cite{SchMos-JCP-90} was used to obtain the TC Hamiltonian.
The $\tcfcihint$ correlation factor is very flexible as it contains explicit electron-nucleus, electron-electron and electron-electron-nucleus terms and have been optimized at the VMC level for each neutral species considered here.
The mean absolute errors (MAE) of the IPs are reported in Table~\ref{tab:mae}.

From Tables~\ref{tab:full-core-a} and \ref{tab:mae} one can observe that the convergence of both the total energies and the ionization potential using  $\tcfcimu$  and $\tcfcihint$ is strongly improved with respect to usual WFT calculations, which is expected due to the presence of explicit correlation.

Several specifics aspects have to be pointed out from these Tables~\ref{tab:full-core-a} and \ref{tab:mae}. \\
\textbf{(i)} For $Z>5$, the total energies provided by $\tcfcimu$ are always closer to the exact ones than that of  $\tcfcihint$. \\
\textbf{(ii)} With increasing nuclear charge, the discrepancy between the total energies at triple-zeta basis set level using $\tcfcimu$ and $\tcfcihint$ increases. This suggests that, the electron-electron-nucleus term of the $\tcfcihint$ takes into account only a part of the correlation effects arising from the core (\textit{i.e.} core-core and core-valence correlation effects). \\
\textbf{(iii)} Although the total energies obtained with  $\tcfcimu$ in double-zeta basis sets can be way below the exact ground state energy (by about 26 mH and 20 mH in the case of the Ne and Ne$^+$ in the cc-pCVDZ basis set, respectively), the energy differences are of good quality (the accuracy of the ionization potential of Ne using  $\tcfcimu$ in the cc-pCVDZ basis set is comparable to that of regular WFT in a cc-pCVQZ basis set).  \\
\textbf{(iv)} While at the double-zeta level the error with respect to the exact IP is significantly smaller using the $\tcfcimu$ than the $\tcfcihint$ approach for all systems, the errors at the triple-zeta level are smaller with the $\tcfcihint$ by approximatively 1 kcal on average. Nevertheless, the error with respect to the exact IP obtained with $\tcfcimu$ at the quadruple-zeta level decrease below 0.001 a.u., showing a systematic convergence pattern.

\section{Conclusion}
\label{sec:conclusion}
In the present work, we further investigated a new strategy based on the TC method which was previously applied on two-electron systems only\cite{Gin-JCP-21}. 
One of the focus of the paper is to test its validity one more realistic systems where many body effects arrise and therefore where the effective three-body terms of the TC Hamiltonian have to be included. 

In order to avoid particle-hole truncation errors due to approximations of the right eigenvector in a given basis set,
the ground state energy of the TC Hamiltonian have been obtained using the recently proposed non-hermitian and three-body variant of the FCIQMC\cite{CohLuoGutDobTewAla-JCP-19}.

The main feature of the recently introduced\cite{Gin-JCP-21}  TC correlation factor is that, beside producing a strictly non divergent TC Hamiltonian, it has a simple parametrization which depends only on a single parameter $\mu$. 
Such a parameter $\mu$ determines the impact of the correlation factor through the depth and typical range of the correlation hole that it induces in the wave function.
Also, thanks to the simple analytical structure of the obtained TC Hamiltonian, all needed integrals can be very efficiently computed in a mixed numerical-analytical scheme. 
The parameter $\mu$ is determined efficiently for each system, 
according to the method described in Ref.~[\citen{Gin-JCP-21}], 
and depends only on the density of the system under study, 
which essentially results in a parameter-free correlation factor.

The main focus of this work is the study of the convergence of the TC eigenvalues and energy differences with respect to the quality of the basis set
and its ability to treat both core and valence electrons.
We performed calculations on the Li--Ne series in the cc-pCVXZ (X=D,T,Q), together with their first cations in order to investigate
the convergence towards the CBS limit of both total energies and IPs.

The main conclusion of this study is that, provided that the basis set contains core-valence functions, very accurate total energies can already be obtained from the triple-zeta quality basis sets.

Regarding the accuracy of the IPs computed here, while the MAE is significantly smaller in double-zeta quality basis sets using the single-parameter $\tcfcimu$ compared to the more elaborate SM-17 correlation factor, 
the results at the triple-zeta level of theory are outperformed by the latter by about 1.2 mH on average.
Nevertheless, the MAE of both these methods are within chemical accuracy (below 1 mH) with CBS limit results at the quadruple zeta level.

In the context of TC calculations, this study shows that the results obtained with a simple one-parameter correlation factor such as $\umu{1}{2}$ are comparable with those obtained with much more sophisticated correlation factors including electron-electron-nucleus terms. 
On the other hand, it should be pointed out that in the TC work of Cohen 
\emph{et al.} the employed SM17 correlation factors were taken from the literature, and were not further optimized for use in the TC method. 
In subsequent work, some of the present authors will investigate a better suited method 
to optimize general Jastrow factors for the TC approach. 

Among the perspective of this work, the use of a correlation factor $\umu{1}{2}$ with a $\mu$ varying in real space could be of interest as it could possibly mimic the electron-electron-nucleus correlation effects. 


\begin{thebibliography}{41}%
	\makeatletter
	\providecommand \@ifxundefined [1]{%
		\@ifx{#1\undefined}
	}%
	\providecommand \@ifnum [1]{%
		\ifnum #1\expandafter \@firstoftwo
		\else \expandafter \@secondoftwo
		\fi
	}%
	\providecommand \@ifx [1]{%
		\ifx #1\expandafter \@firstoftwo
		\else \expandafter \@secondoftwo
		\fi
	}%
	\providecommand \natexlab [1]{#1}%
	\providecommand \enquote  [1]{``#1''}%
	\providecommand \bibnamefont  [1]{#1}%
	\providecommand \bibfnamefont [1]{#1}%
	\providecommand \citenamefont [1]{#1}%
	\providecommand \href@noop [0]{\@secondoftwo}%
	\providecommand \href [0]{\begingroup \@sanitize@url \@href}%
	\providecommand \@href[1]{\@@startlink{#1}\@@href}%
	\providecommand \@@href[1]{\endgroup#1\@@endlink}%
	\providecommand \@sanitize@url [0]{\catcode `\\12\catcode `\$12\catcode
		`\&12\catcode `\#12\catcode `\^12\catcode `\_12\catcode `\%12\relax}%
	\providecommand \@@startlink[1]{}%
	\providecommand \@@endlink[0]{}%
	\providecommand \url  [0]{\begingroup\@sanitize@url \@url }%
	\providecommand \@url [1]{\endgroup\@href {#1}{\urlprefix }}%
	\providecommand \urlprefix  [0]{URL }%
	\providecommand \Eprint [0]{\href }%
	\providecommand \doibase [0]{http://dx.doi.org/}%
	\providecommand \selectlanguage [0]{\@gobble}%
	\providecommand \bibinfo  [0]{\@secondoftwo}%
	\providecommand \bibfield  [0]{\@secondoftwo}%
	\providecommand \translation [1]{[#1]}%
	\providecommand \BibitemOpen [0]{}%
	\providecommand \bibitemStop [0]{}%
	\providecommand \bibitemNoStop [0]{.\EOS\space}%
	\providecommand \EOS [0]{\spacefactor3000\relax}%
	\providecommand \BibitemShut  [1]{\csname bibitem#1\endcsname}%
	\let\auto@bib@innerbib\@empty
	\bibitem [{\citenamefont {Kato}(1957)}]{Kat-CPAM-57}%
	\BibitemOpen
	\bibfield  {author} {\bibinfo {author} {\bibfnamefont {T.}~\bibnamefont
			{Kato}},\ }\href {\doibase 10.1002/cpa.3160100201} {\bibfield  {journal}
		{\bibinfo  {journal} {Comm. Pure Appl. Math.}\ }\textbf {\bibinfo {volume}
			{10}},\ \bibinfo {pages} {151} (\bibinfo {year} {1957})}\BibitemShut
	{NoStop}%
	\bibitem [{\citenamefont {Ten-no}(2012)}]{Ten-TCA-12}%
	\BibitemOpen
	\bibfield  {author} {\bibinfo {author} {\bibfnamefont {S.}~\bibnamefont
			{Ten-no}},\ }\href {\doibase 10.1007/s00214-011-1070-1} {\bibfield  {journal}
		{\bibinfo  {journal} {Theor. Chem. Acc.}\ }\textbf {\bibinfo {volume}
			{131}},\ \bibinfo {pages} {1070} (\bibinfo {year} {2012})}\BibitemShut
	{NoStop}%
	\bibitem [{\citenamefont {Ten-no}\ and\ \citenamefont
		{Noga}(2012)}]{TenNog-WIREs-12}%
	\BibitemOpen
	\bibfield  {author} {\bibinfo {author} {\bibfnamefont {S.}~\bibnamefont
			{Ten-no}}\ and\ \bibinfo {author} {\bibfnamefont {J.}~\bibnamefont {Noga}},\
	}\href {\doibase 10.1002/wcms.68} {\bibfield  {journal} {\bibinfo  {journal}
			{WIREs Comput. Mol. Sci.}\ }\textbf {\bibinfo {volume} {2}},\ \bibinfo
		{pages} {114} (\bibinfo {year} {2012})}\BibitemShut {NoStop}%
	\bibitem [{\citenamefont {Hattig}\ \emph {et~al.}(2012)\citenamefont {Hattig},
		\citenamefont {Klopper}, \citenamefont {Kohn},\ and\ \citenamefont
		{Tew}}]{HatKloKohTew-CR-12}%
	\BibitemOpen
	\bibfield  {author} {\bibinfo {author} {\bibfnamefont {C.}~\bibnamefont
			{Hattig}}, \bibinfo {author} {\bibfnamefont {W.}~\bibnamefont {Klopper}},
		\bibinfo {author} {\bibfnamefont {A.}~\bibnamefont {Kohn}}, \ and\ \bibinfo
		{author} {\bibfnamefont {D.~P.}\ \bibnamefont {Tew}},\ }\href {\doibase
		10.1021/cr200168z} {\bibfield  {journal} {\bibinfo  {journal} {Chem. Rev.}\
		}\textbf {\bibinfo {volume} {112}},\ \bibinfo {pages} {4} (\bibinfo {year}
		{2012})}\BibitemShut {NoStop}%
	\bibitem [{\citenamefont {Kong}, \citenamefont {Bischo},\ and\ \citenamefont
		{Valeev}(2012)}]{KonBisVal-CR-12}%
	\BibitemOpen
	\bibfield  {author} {\bibinfo {author} {\bibfnamefont {L.}~\bibnamefont
			{Kong}}, \bibinfo {author} {\bibfnamefont {F.~A.}\ \bibnamefont {Bischo}}, \
		and\ \bibinfo {author} {\bibfnamefont {E.~F.}\ \bibnamefont {Valeev}},\
	}\href {\doibase 10.1021/cr200204r} {\bibfield  {journal} {\bibinfo
			{journal} {Chem. Rev.}\ }\textbf {\bibinfo {volume} {112}},\ \bibinfo {pages}
		{75} (\bibinfo {year} {2012})}\BibitemShut {NoStop}%
	\bibitem [{\citenamefont {Gr\"uneis}\ \emph {et~al.}(2017)\citenamefont
		{Gr\"uneis}, \citenamefont {Hirata}, \citenamefont {Ohnishi},\ and\
		\citenamefont {Ten-no}}]{GruHirOhnTen-JCP-17}%
	\BibitemOpen
	\bibfield  {author} {\bibinfo {author} {\bibfnamefont {A.}~\bibnamefont
			{Gr\"uneis}}, \bibinfo {author} {\bibfnamefont {S.}~\bibnamefont {Hirata}},
		\bibinfo {author} {\bibfnamefont {Y.-Y.}\ \bibnamefont {Ohnishi}}, \ and\
		\bibinfo {author} {\bibfnamefont {S.}~\bibnamefont {Ten-no}},\ }\href
	{\doibase 10.1063/1.4976974} {\bibfield  {journal} {\bibinfo  {journal} {J.
				Chem. Phys.}\ }\textbf {\bibinfo {volume} {146}},\ \bibinfo {pages} {080901}
		(\bibinfo {year} {2017})}\BibitemShut {NoStop}%
	\bibitem [{\citenamefont {Ma}\ and\ \citenamefont
		{Werner}(2018)}]{MaWer-WIREs-18}%
	\BibitemOpen
	\bibfield  {author} {\bibinfo {author} {\bibfnamefont {Q.}~\bibnamefont
			{Ma}}\ and\ \bibinfo {author} {\bibfnamefont {H.-J.}\ \bibnamefont
			{Werner}},\ }\href {\doibase 10.1002/wcms.1371} {\bibfield  {journal}
		{\bibinfo  {journal} {WIREs Comput. Mol. Sci.}\ }\textbf {\bibinfo {volume}
			{8}},\ \bibinfo {pages} {e1371} (\bibinfo {year} {2018})}\BibitemShut
	{NoStop}%
	\bibitem [{\citenamefont {Toulouse}, \citenamefont {Assaraf},\ and\
		\citenamefont {Umrigar}(2016)}]{TouAssUmr-book-vmc}%
	\BibitemOpen
	\bibfield  {author} {\bibinfo {author} {\bibfnamefont {J.}~\bibnamefont
			{Toulouse}}, \bibinfo {author} {\bibfnamefont {R.}~\bibnamefont {Assaraf}}, \
		and\ \bibinfo {author} {\bibfnamefont {C.~J.}\ \bibnamefont {Umrigar}},\ }in\
	\href {\doibase 10.1016/bs.aiq.2015.07.003} {\emph {\bibinfo {booktitle}
			{{Advances in Quantum Chemistry}}}},\ Vol.~\bibinfo {volume} {73}\ (\bibinfo
	{publisher} {Academic Press},\ \bibinfo {address} {Cambridge, MA, USA},\
	\bibinfo {year} {2016})\ pp.\ \bibinfo {pages} {285--314}\BibitemShut
	{NoStop}%
	\bibitem [{\citenamefont {Hirschfelder}(1963)}]{Hirschfelder-JCP-63}%
	\BibitemOpen
	\bibfield  {author} {\bibinfo {author} {\bibfnamefont {J.~O.}\ \bibnamefont
			{Hirschfelder}},\ }\href {\doibase 10.1063/1.1734157} {\bibfield  {journal}
		{\bibinfo  {journal} {The Journal of Chemical Physics}\ }\textbf {\bibinfo
			{volume} {39}},\ \bibinfo {pages} {3145} (\bibinfo {year}
		{1963})}\BibitemShut {NoStop}%
	\bibitem [{\citenamefont {Boys}\ and\ \citenamefont
		{Handy}(1969{\natexlab{a}})}]{BoyHan-PRSLA-69}%
	\BibitemOpen
	\bibfield  {author} {\bibinfo {author} {\bibfnamefont {F.~S.}\ \bibnamefont
			{Boys}}\ and\ \bibinfo {author} {\bibnamefont {Handy}},\ }\href {\doibase
		10.1098/rspa.1969.0061} {\bibfield  {journal} {\bibinfo  {journal} {Proc. R.
				Soc. Lond. A.}\ }\textbf {\bibinfo {volume} {310}},\ \bibinfo {pages} {43}
		(\bibinfo {year} {1969}{\natexlab{a}})}\BibitemShut {NoStop}%
	\bibitem [{\citenamefont {Boys}\ and\ \citenamefont
		{Handy}(1969{\natexlab{b}})}]{BoyHan-2-PRSLA-69}%
	\BibitemOpen
	\bibfield  {author} {\bibinfo {author} {\bibfnamefont {F.~S.}\ \bibnamefont
			{Boys}}\ and\ \bibinfo {author} {\bibfnamefont {C.~N.}\ \bibnamefont
			{Handy}},\ }\href {\doibase 10.1098/rspa.1969.0062} {\bibfield  {journal}
		{\bibinfo  {journal} {Proc. R. Soc. Lond. A.}\ }\textbf {\bibinfo {volume}
			{310}},\ \bibinfo {pages} {63} (\bibinfo {year}
		{1969}{\natexlab{b}})}\BibitemShut {NoStop}%
	\bibitem [{\citenamefont {Ten-no}(2000)}]{TenNo-CPL-00-a}%
	\BibitemOpen
	\bibfield  {author} {\bibinfo {author} {\bibfnamefont {S.}~\bibnamefont
			{Ten-no}},\ }\href {\doibase https://doi.org/10.1016/S0009-2614(00)01066-6}
	{\bibfield  {journal} {\bibinfo  {journal} {Chemical Physics Letters}\
		}\textbf {\bibinfo {volume} {330}},\ \bibinfo {pages} {169 } (\bibinfo {year}
		{2000})}\BibitemShut {NoStop}%
	\bibitem [{\citenamefont {Hino}, \citenamefont {Tanimura},\ and\ \citenamefont
		{Ten-no}(2001)}]{HinTanTen-JCP-01}%
	\BibitemOpen
	\bibfield  {author} {\bibinfo {author} {\bibfnamefont {O.}~\bibnamefont
			{Hino}}, \bibinfo {author} {\bibfnamefont {Y.}~\bibnamefont {Tanimura}}, \
		and\ \bibinfo {author} {\bibfnamefont {S.}~\bibnamefont {Ten-no}},\ }\href
	{\doibase 10.1063/1.1408299} {\bibfield  {journal} {\bibinfo  {journal} {The
				Journal of Chemical Physics}\ }\textbf {\bibinfo {volume} {115}},\ \bibinfo
		{pages} {7865} (\bibinfo {year} {2001})},\ \Eprint
	{http://arxiv.org/abs/https://doi.org/10.1063/1.1408299}
	{https://doi.org/10.1063/1.1408299} \BibitemShut {NoStop}%
	\bibitem [{\citenamefont {Hino}, \citenamefont {Tanimura},\ and\ \citenamefont
		{Ten-no}(2002)}]{HinTanTen-CPL-02}%
	\BibitemOpen
	\bibfield  {author} {\bibinfo {author} {\bibfnamefont {O.}~\bibnamefont
			{Hino}}, \bibinfo {author} {\bibfnamefont {Y.}~\bibnamefont {Tanimura}}, \
		and\ \bibinfo {author} {\bibfnamefont {S.}~\bibnamefont {Ten-no}},\ }\href
	{\doibase https://doi.org/10.1016/S0009-2614(02)00042-8} {\bibfield
		{journal} {\bibinfo  {journal} {Chemical Physics Letters}\ }\textbf {\bibinfo
			{volume} {353}},\ \bibinfo {pages} {317 } (\bibinfo {year}
		{2002})}\BibitemShut {NoStop}%
	\bibitem [{\citenamefont {Cohen}\ \emph {et~al.}(2019)\citenamefont {Cohen},
		\citenamefont {Luo}, \citenamefont {Guther}, \citenamefont {Dobrautz},
		\citenamefont {Tew},\ and\ \citenamefont
		{Alavi}}]{CohLuoGutDobTewAla-JCP-19}%
	\BibitemOpen
	\bibfield  {author} {\bibinfo {author} {\bibfnamefont {A.~J.}\ \bibnamefont
			{Cohen}}, \bibinfo {author} {\bibfnamefont {H.}~\bibnamefont {Luo}}, \bibinfo
		{author} {\bibfnamefont {K.}~\bibnamefont {Guther}}, \bibinfo {author}
		{\bibfnamefont {W.}~\bibnamefont {Dobrautz}}, \bibinfo {author}
		{\bibfnamefont {D.~P.}\ \bibnamefont {Tew}}, \ and\ \bibinfo {author}
		{\bibfnamefont {A.}~\bibnamefont {Alavi}},\ }\href {\doibase
		10.1063/1.5116024} {\bibfield  {journal} {\bibinfo  {journal} {The Journal of
				Chemical Physics}\ }\textbf {\bibinfo {volume} {151}},\ \bibinfo {pages}
		{061101} (\bibinfo {year} {2019})},\ \Eprint
	{http://arxiv.org/abs/https://doi.org/10.1063/1.5116024}
	{https://doi.org/10.1063/1.5116024} \BibitemShut {NoStop}%
	\bibitem [{\citenamefont {Schmidt}\ and\ \citenamefont
		{Moskowitz}(1990)}]{SchMos-JCP-90}%
	\BibitemOpen
	\bibfield  {author} {\bibinfo {author} {\bibfnamefont {K.~E.}\ \bibnamefont
			{Schmidt}}\ and\ \bibinfo {author} {\bibfnamefont {J.~W.}\ \bibnamefont
			{Moskowitz}},\ }\href {\doibase 10.1063/1.458750} {\bibfield  {journal}
		{\bibinfo  {journal} {The Journal of Chemical Physics}\ }\textbf {\bibinfo
			{volume} {93}},\ \bibinfo {pages} {4172} (\bibinfo {year} {1990})},\ \Eprint
	{http://arxiv.org/abs/https://doi.org/10.1063/1.458750}
	{https://doi.org/10.1063/1.458750} \BibitemShut {NoStop}%
	\bibitem [{\citenamefont {Dobrautz}, \citenamefont {Luo},\ and\ \citenamefont
		{Alavi}(2019)}]{DobLuoAla-PRB-19}%
	\BibitemOpen
	\bibfield  {author} {\bibinfo {author} {\bibfnamefont {W.}~\bibnamefont
			{Dobrautz}}, \bibinfo {author} {\bibfnamefont {H.}~\bibnamefont {Luo}}, \
		and\ \bibinfo {author} {\bibfnamefont {A.}~\bibnamefont {Alavi}},\ }\href
	{\doibase 10.1103/PhysRevB.99.075119} {\bibfield  {journal} {\bibinfo
			{journal} {Phys. Rev. B}\ }\textbf {\bibinfo {volume} {99}},\ \bibinfo
		{pages} {075119} (\bibinfo {year} {2019})}\BibitemShut {NoStop}%
	\bibitem [{\citenamefont {Gutzwiller}(1963)}]{Gutzwiller-PRL-63}%
	\BibitemOpen
	\bibfield  {author} {\bibinfo {author} {\bibfnamefont {M.~C.}\ \bibnamefont
			{Gutzwiller}},\ }\href {\doibase 10.1103/PhysRevLett.10.159} {\bibfield
		{journal} {\bibinfo  {journal} {Phys. Rev. Lett.}\ }\textbf {\bibinfo
			{volume} {10}},\ \bibinfo {pages} {159} (\bibinfo {year} {1963})}\BibitemShut
	{NoStop}%
	\bibitem [{\citenamefont {Brinkman}\ and\ \citenamefont
		{Rice}(1970)}]{BriRic-PRB-70}%
	\BibitemOpen
	\bibfield  {author} {\bibinfo {author} {\bibfnamefont {W.~F.}\ \bibnamefont
			{Brinkman}}\ and\ \bibinfo {author} {\bibfnamefont {T.~M.}\ \bibnamefont
			{Rice}},\ }\href {\doibase 10.1103/PhysRevB.2.4302} {\bibfield  {journal}
		{\bibinfo  {journal} {Phys. Rev. B}\ }\textbf {\bibinfo {volume} {2}},\
		\bibinfo {pages} {4302} (\bibinfo {year} {1970})}\BibitemShut {NoStop}%
	\bibitem [{\citenamefont {Baiardi}\ and\ \citenamefont
		{Reiher}(2020)}]{BaiRei-JCP-20}%
	\BibitemOpen
	\bibfield  {author} {\bibinfo {author} {\bibfnamefont {A.}~\bibnamefont
			{Baiardi}}\ and\ \bibinfo {author} {\bibfnamefont {M.}~\bibnamefont
			{Reiher}},\ }\href {\doibase 10.1063/1.5129672} {\bibfield  {journal}
		{\bibinfo  {journal} {The Journal of Chemical Physics}\ }\textbf {\bibinfo
			{volume} {152}},\ \bibinfo {pages} {040903} (\bibinfo {year} {2020})},\
	\Eprint {http://arxiv.org/abs/https://doi.org/10.1063/1.5129672}
	{https://doi.org/10.1063/1.5129672} \BibitemShut {NoStop}%
	\bibitem [{\citenamefont {Motta}\ \emph {et~al.}(2020)\citenamefont {Motta},
		\citenamefont {Gujarati}, \citenamefont {Rice}, \citenamefont {Kumar},
		\citenamefont {Masteran}, \citenamefont {Latone}, \citenamefont {Lee},
		\citenamefont {Valeev},\ and\ \citenamefont {Takeshita}}]{ValTak-PCCP-20}%
	\BibitemOpen
	\bibfield  {author} {\bibinfo {author} {\bibfnamefont {M.}~\bibnamefont
			{Motta}}, \bibinfo {author} {\bibfnamefont {T.~P.}\ \bibnamefont {Gujarati}},
		\bibinfo {author} {\bibfnamefont {J.~E.}\ \bibnamefont {Rice}}, \bibinfo
		{author} {\bibfnamefont {A.}~\bibnamefont {Kumar}}, \bibinfo {author}
		{\bibfnamefont {C.}~\bibnamefont {Masteran}}, \bibinfo {author}
		{\bibfnamefont {J.~A.}\ \bibnamefont {Latone}}, \bibinfo {author}
		{\bibfnamefont {E.}~\bibnamefont {Lee}}, \bibinfo {author} {\bibfnamefont
			{E.~F.}\ \bibnamefont {Valeev}}, \ and\ \bibinfo {author} {\bibfnamefont
			{T.~Y.}\ \bibnamefont {Takeshita}},\ }\href {\doibase 10.1039/D0CP04106H}
	{\bibfield  {journal} {\bibinfo  {journal} {Phys. Chem. Chem. Phys.}\
		}\textbf {\bibinfo {volume} {22}},\ \bibinfo {pages} {24270} (\bibinfo {year}
		{2020})}\BibitemShut {NoStop}%
	\bibitem [{\citenamefont {McArdle}\ and\ \citenamefont
		{Tew}(2020)}]{McArdle2020}%
	\BibitemOpen
	\bibfield  {author} {\bibinfo {author} {\bibfnamefont {S.}~\bibnamefont
			{McArdle}}\ and\ \bibinfo {author} {\bibfnamefont {D.~P.}\ \bibnamefont
			{Tew}},\ }\href@noop {} {\enquote {\bibinfo {title} {Improving the accuracy
				of quantum computational chemistry using the transcorrelated method},}\ }
	(\bibinfo {year} {2020}),\ \bibinfo {note}
	{\url{https://arxiv.org/abs/2006.11181}},\ \Eprint
	{http://arxiv.org/abs/arXiv:2006.11181} {arXiv:2006.11181} \BibitemShut
	{NoStop}%
	\bibitem [{\citenamefont {Schleich}, \citenamefont {Kottmann},\ and\
		\citenamefont {Aspuru-Guzik}(2021)}]{Schleich2021}%
	\BibitemOpen
	\bibfield  {author} {\bibinfo {author} {\bibfnamefont {P.}~\bibnamefont
			{Schleich}}, \bibinfo {author} {\bibfnamefont {J.~S.}\ \bibnamefont
			{Kottmann}}, \ and\ \bibinfo {author} {\bibfnamefont {A.}~\bibnamefont
			{Aspuru-Guzik}},\ }\href@noop {} {\enquote {\bibinfo {title} {Improving the
				accuracy of the variational quantum eigensolver for molecular systems by the
				explicitly-correlated perturbative [2]$_\text{R12}$-correction},}\ }
	(\bibinfo {year} {2021}),\ \bibinfo {note}
	{\url{https://arxiv.org/abs/2110.06812}},\ \Eprint
	{http://arxiv.org/abs/2110.06812} {arXiv:2110.06812} \BibitemShut {NoStop}%
	\bibitem [{\citenamefont {Kumar}\ \emph {et~al.}(2022)\citenamefont {Kumar},
		\citenamefont {Asthana}, \citenamefont {Masteran}, \citenamefont {Valeev},
		\citenamefont {Zhang}, \citenamefont {Cincio}, \citenamefont {Tretiak},\ and\
		\citenamefont {Dub}}]{Kumar2022}%
	\BibitemOpen
	\bibfield  {author} {\bibinfo {author} {\bibfnamefont {A.}~\bibnamefont
			{Kumar}}, \bibinfo {author} {\bibfnamefont {A.}~\bibnamefont {Asthana}},
		\bibinfo {author} {\bibfnamefont {C.}~\bibnamefont {Masteran}}, \bibinfo
		{author} {\bibfnamefont {E.~F.}\ \bibnamefont {Valeev}}, \bibinfo {author}
		{\bibfnamefont {Y.}~\bibnamefont {Zhang}}, \bibinfo {author} {\bibfnamefont
			{L.}~\bibnamefont {Cincio}}, \bibinfo {author} {\bibfnamefont
			{S.}~\bibnamefont {Tretiak}}, \ and\ \bibinfo {author} {\bibfnamefont
			{P.~A.}\ \bibnamefont {Dub}},\ }\href@noop {} {\enquote {\bibinfo {title}
			{Accurate quantum simulation of molecular ground and excited states with a
				transcorrelated hamiltonian},}\ } (\bibinfo {year} {2022}),\ \bibinfo {note}
	{\url{https://arxiv.org/abs/2201.09852}},\ \Eprint
	{http://arxiv.org/abs/2201.09852} {arXiv:2201.09852} \BibitemShut {NoStop}%
	\bibitem [{\citenamefont {Sokolov}\ \emph {et~al.}(2022)\citenamefont
		{Sokolov}, \citenamefont {Dobrautz}, \citenamefont {Luo}, \citenamefont
		{Alavi},\ and\ \citenamefont {Tavernelli}}]{Sokolov2022}%
	\BibitemOpen
	\bibfield  {author} {\bibinfo {author} {\bibfnamefont {I.~O.}\ \bibnamefont
			{Sokolov}}, \bibinfo {author} {\bibfnamefont {W.}~\bibnamefont {Dobrautz}},
		\bibinfo {author} {\bibfnamefont {H.}~\bibnamefont {Luo}}, \bibinfo {author}
		{\bibfnamefont {A.}~\bibnamefont {Alavi}}, \ and\ \bibinfo {author}
		{\bibfnamefont {I.}~\bibnamefont {Tavernelli}},\ }\href@noop {} {\enquote
		{\bibinfo {title} {Orders of magnitude reduction in the computational
				overhead for quantum many-body problems on quantum computers via an exact
				transcorrelated method},}\ } (\bibinfo {year} {2022}),\ \bibinfo {note}
	{\url{https://arxiv.org/abs/2201.03049}},\ \Eprint
	{http://arxiv.org/abs/2201.03049} {arXiv:2201.03049} \BibitemShut {NoStop}%
	\bibitem [{\citenamefont {Giner}(2021)}]{Gin-JCP-21}%
	\BibitemOpen
	\bibfield  {author} {\bibinfo {author} {\bibfnamefont {E.}~\bibnamefont
			{Giner}},\ }\href {\doibase 10.1063/5.0044683} {\bibfield  {journal}
		{\bibinfo  {journal} {J. Chem. Phys.}\ }\textbf {\bibinfo {volume} {154}},\
		\bibinfo {pages} {084119} (\bibinfo {year} {2021})}\BibitemShut {NoStop}%
	\bibitem [{\citenamefont {Guther}\ \emph {et~al.}(2021)\citenamefont {Guther},
		\citenamefont {Cohen}, \citenamefont {Luo},\ and\ \citenamefont
		{Alavi}}]{GutCohLuoAla-JCP-21}%
	\BibitemOpen
	\bibfield  {author} {\bibinfo {author} {\bibfnamefont {K.}~\bibnamefont
			{Guther}}, \bibinfo {author} {\bibfnamefont {A.~J.}\ \bibnamefont {Cohen}},
		\bibinfo {author} {\bibfnamefont {H.}~\bibnamefont {Luo}}, \ and\ \bibinfo
		{author} {\bibfnamefont {A.}~\bibnamefont {Alavi}},\ }\href {\doibase
		10.1063/5.0055575} {\bibfield  {journal} {\bibinfo  {journal} {J. Chem.
				Phys.}\ }\textbf {\bibinfo {volume} {155}},\ \bibinfo {pages} {011102}
		(\bibinfo {year} {2021})}\BibitemShut {NoStop}%
	\bibitem [{\citenamefont {Booth}, \citenamefont {Thom},\ and\ \citenamefont
		{Alavi}(2009)}]{BooThoAla-JCP-09}%
	\BibitemOpen
	\bibfield  {author} {\bibinfo {author} {\bibfnamefont {G.~H.}\ \bibnamefont
			{Booth}}, \bibinfo {author} {\bibfnamefont {A.~J.~W.}\ \bibnamefont {Thom}},
		\ and\ \bibinfo {author} {\bibfnamefont {A.}~\bibnamefont {Alavi}},\ }\href
	{https://doi.org/10.1063/1.3193710} {\bibfield  {journal} {\bibinfo
			{journal} {J. Chem. Phys.}\ }\textbf {\bibinfo {volume} {{131}}} (\bibinfo
		{year} {{2009}})}\BibitemShut {NoStop}%
	\bibitem [{\citenamefont {Booth}\ and\ \citenamefont
		{Alavi}(2010)}]{BooAla-JCP-10}%
	\BibitemOpen
	\bibfield  {author} {\bibinfo {author} {\bibfnamefont {G.~H.}\ \bibnamefont
			{Booth}}\ and\ \bibinfo {author} {\bibfnamefont {A.}~\bibnamefont {Alavi}},\
	}\href {\doibase 10.1063/1.3407895} {\bibfield  {journal} {\bibinfo
			{journal} {J. Chem. Phys.}\ }\textbf {\bibinfo {volume} {132}},\ \bibinfo
		{pages} {174104} (\bibinfo {year} {2010})}\BibitemShut {NoStop}%
	\bibitem [{\citenamefont {Booth}\ \emph {et~al.}(2011)\citenamefont {Booth},
		\citenamefont {Cleland}, \citenamefont {Thom},\ and\ \citenamefont
		{Alavi}}]{BooCleThoAla-JCP-11}%
	\BibitemOpen
	\bibfield  {author} {\bibinfo {author} {\bibfnamefont {G.~H.}\ \bibnamefont
			{Booth}}, \bibinfo {author} {\bibfnamefont {D.}~\bibnamefont {Cleland}},
		\bibinfo {author} {\bibfnamefont {A.~J.~W.}\ \bibnamefont {Thom}}, \ and\
		\bibinfo {author} {\bibfnamefont {A.}~\bibnamefont {Alavi}},\ }\href
	{https://doi.org/10.1063/1.3624383} {\bibfield  {journal} {\bibinfo
			{journal} {J. Chem. Phys.}\ }\textbf {\bibinfo {volume} {135}},\ \bibinfo
		{pages} {084104} (\bibinfo {year} {2011})}\BibitemShut {NoStop}%
	\bibitem [{\citenamefont {Ghanem}, \citenamefont {Lozovoi},\ and\ \citenamefont
		{Alavi}(2019)}]{GhaLozAla-JCP-19}%
	\BibitemOpen
	\bibfield  {author} {\bibinfo {author} {\bibfnamefont {K.}~\bibnamefont
			{Ghanem}}, \bibinfo {author} {\bibfnamefont {A.~Y.}\ \bibnamefont {Lozovoi}},
		\ and\ \bibinfo {author} {\bibfnamefont {A.}~\bibnamefont {Alavi}},\ }\href
	{\doibase 10.1063/1.5134006} {\bibfield  {journal} {\bibinfo  {journal} {The
				Journal of Chemical Physics}\ }\textbf {\bibinfo {volume} {151}},\ \bibinfo
		{pages} {224108} (\bibinfo {year} {2019})},\ \Eprint
	{http://arxiv.org/abs/https://doi.org/10.1063/1.5134006}
	{https://doi.org/10.1063/1.5134006} \BibitemShut {NoStop}%
	\bibitem [{\citenamefont {Vitale}, \citenamefont {Alavi},\ and\ \citenamefont
		{Kats}(2020)}]{VitAlaKat-JCTC-20}%
	\BibitemOpen
	\bibfield  {author} {\bibinfo {author} {\bibfnamefont {E.}~\bibnamefont
			{Vitale}}, \bibinfo {author} {\bibfnamefont {A.}~\bibnamefont {Alavi}}, \
		and\ \bibinfo {author} {\bibfnamefont {D.}~\bibnamefont {Kats}},\ }\href
	{\doibase 10.1021/acs.jctc.0c00470} {\bibfield  {journal} {\bibinfo
			{journal} {Journal of Chemical Theory and Computation}\ }\textbf {\bibinfo
			{volume} {16}},\ \bibinfo {pages} {5621} (\bibinfo {year} {2020})},\ \bibinfo
	{note} {pMID: 32786911},\ \Eprint
	{http://arxiv.org/abs/https://doi.org/10.1021/acs.jctc.0c00470}
	{https://doi.org/10.1021/acs.jctc.0c00470} \BibitemShut {NoStop}%
	\bibitem [{\citenamefont {Guther}\ \emph {et~al.}(2020)\citenamefont {Guther},
		\citenamefont {Anderson}, \citenamefont {Blunt}, \citenamefont {Bogdanov},
		\citenamefont {Cleland}, \citenamefont {Dattani}, \citenamefont {Dobrautz},
		\citenamefont {Ghanem}, \citenamefont {Jeszenszki}, \citenamefont
		{Liebermann}, \citenamefont {Manni}, \citenamefont {Lozovoi}, \citenamefont
		{Luo}, \citenamefont {Ma}, \citenamefont {Merz}, \citenamefont {Overy},
		\citenamefont {Rampp}, \citenamefont {Samanta}, \citenamefont {Schwarz},
		\citenamefont {Shepherd}, \citenamefont {Smart}, \citenamefont {Vitale},
		\citenamefont {Weser}, \citenamefont {Booth},\ and\ \citenamefont
		{Alavi}}]{Guther2020}%
	\BibitemOpen
	\bibfield  {author} {\bibinfo {author} {\bibfnamefont {K.}~\bibnamefont
			{Guther}}, \bibinfo {author} {\bibfnamefont {R.~J.}\ \bibnamefont
			{Anderson}}, \bibinfo {author} {\bibfnamefont {N.~S.}\ \bibnamefont {Blunt}},
		\bibinfo {author} {\bibfnamefont {N.~A.}\ \bibnamefont {Bogdanov}}, \bibinfo
		{author} {\bibfnamefont {D.}~\bibnamefont {Cleland}}, \bibinfo {author}
		{\bibfnamefont {N.}~\bibnamefont {Dattani}}, \bibinfo {author} {\bibfnamefont
			{W.}~\bibnamefont {Dobrautz}}, \bibinfo {author} {\bibfnamefont
			{K.}~\bibnamefont {Ghanem}}, \bibinfo {author} {\bibfnamefont
			{P.}~\bibnamefont {Jeszenszki}}, \bibinfo {author} {\bibfnamefont
			{N.}~\bibnamefont {Liebermann}}, \bibinfo {author} {\bibfnamefont {G.~L.}\
			\bibnamefont {Manni}}, \bibinfo {author} {\bibfnamefont {A.~Y.}\ \bibnamefont
			{Lozovoi}}, \bibinfo {author} {\bibfnamefont {H.}~\bibnamefont {Luo}},
		\bibinfo {author} {\bibfnamefont {D.}~\bibnamefont {Ma}}, \bibinfo {author}
		{\bibfnamefont {F.}~\bibnamefont {Merz}}, \bibinfo {author} {\bibfnamefont
			{C.}~\bibnamefont {Overy}}, \bibinfo {author} {\bibfnamefont
			{M.}~\bibnamefont {Rampp}}, \bibinfo {author} {\bibfnamefont {P.~K.}\
			\bibnamefont {Samanta}}, \bibinfo {author} {\bibfnamefont {L.~R.}\
			\bibnamefont {Schwarz}}, \bibinfo {author} {\bibfnamefont {J.~J.}\
			\bibnamefont {Shepherd}}, \bibinfo {author} {\bibfnamefont {S.~D.}\
			\bibnamefont {Smart}}, \bibinfo {author} {\bibfnamefont {E.}~\bibnamefont
			{Vitale}}, \bibinfo {author} {\bibfnamefont {O.}~\bibnamefont {Weser}},
		\bibinfo {author} {\bibfnamefont {G.~H.}\ \bibnamefont {Booth}}, \ and\
		\bibinfo {author} {\bibfnamefont {A.}~\bibnamefont {Alavi}},\ }\href
	{\doibase 10.1063/5.0005754} {\bibfield  {journal} {\bibinfo  {journal} {The
				Journal of Chemical Physics}\ }\textbf {\bibinfo {volume} {153}},\ \bibinfo
		{pages} {034107} (\bibinfo {year} {2020})}\BibitemShut {NoStop}%
	\bibitem [{\citenamefont {Dobrautz}, \citenamefont {Smart},\ and\ \citenamefont
		{Alavi}(2019)}]{Dobrautz2019}%
	\BibitemOpen
	\bibfield  {author} {\bibinfo {author} {\bibfnamefont {W.}~\bibnamefont
			{Dobrautz}}, \bibinfo {author} {\bibfnamefont {S.~D.}\ \bibnamefont {Smart}},
		\ and\ \bibinfo {author} {\bibfnamefont {A.}~\bibnamefont {Alavi}},\ }\href
	{\doibase 10.1063/1.5108908} {\bibfield  {journal} {\bibinfo  {journal} {The
				Journal of Chemical Physics}\ }\textbf {\bibinfo {volume} {151}},\ \bibinfo
		{pages} {094104} (\bibinfo {year} {2019})}\BibitemShut {NoStop}%
	\bibitem [{\citenamefont {Dobrautz}\ \emph {et~al.}(2021)\citenamefont
		{Dobrautz}, \citenamefont {Weser}, \citenamefont {Bogdanov}, \citenamefont
		{Alavi},\ and\ \citenamefont {Li~Manni}}]{Dobrautz2021}%
	\BibitemOpen
	\bibfield  {author} {\bibinfo {author} {\bibfnamefont {W.}~\bibnamefont
			{Dobrautz}}, \bibinfo {author} {\bibfnamefont {O.}~\bibnamefont {Weser}},
		\bibinfo {author} {\bibfnamefont {N.~A.}\ \bibnamefont {Bogdanov}}, \bibinfo
		{author} {\bibfnamefont {A.}~\bibnamefont {Alavi}}, \ and\ \bibinfo {author}
		{\bibfnamefont {G.}~\bibnamefont {Li~Manni}},\ }\href {\doibase
		10.1021/acs.jctc.1c00589} {\bibfield  {journal} {\bibinfo  {journal} {Journal
				of Chemical Theory and Computation}\ }\textbf {\bibinfo {volume} {17}},\
		\bibinfo {pages} {5684} (\bibinfo {year} {2021})},\ \bibinfo {note} {pMID:
		34469685}\BibitemShut {NoStop}%
	\bibitem [{\citenamefont {Becke}(1988)}]{Bec-JCP-88b}%
	\BibitemOpen
	\bibfield  {author} {\bibinfo {author} {\bibfnamefont {A.~D.}\ \bibnamefont
			{Becke}},\ }\href@noop {} {\bibfield  {journal} {\bibinfo  {journal} {J.
				Chem. Phys.}\ }\textbf {\bibinfo {volume} {{88}}},\ \bibinfo {pages} {2547}
		(\bibinfo {year} {1988})}\BibitemShut {NoStop}%
	\bibitem [{\citenamefont {Huron}, \citenamefont {Malrieu},\ and\ \citenamefont
		{Rancurel}(1973)}]{malrieu_cipsi}%
	\BibitemOpen
	\bibfield  {author} {\bibinfo {author} {\bibfnamefont {B.}~\bibnamefont
			{Huron}}, \bibinfo {author} {\bibfnamefont {J.~P.}\ \bibnamefont {Malrieu}},
		\ and\ \bibinfo {author} {\bibfnamefont {P.}~\bibnamefont {Rancurel}},\
	}\href@noop {} {\bibfield  {journal} {\bibinfo  {journal} {J. Chem. Phys.}\
		}\textbf {\bibinfo {volume} {58}},\ \bibinfo {pages} {5745} (\bibinfo {year}
		{1973})}\BibitemShut {NoStop}%
	\bibitem [{\citenamefont {Garniron}\ \emph {et~al.}(2019)\citenamefont
		{Garniron}, \citenamefont {Gasperich}, \citenamefont {Applencourt},
		\citenamefont {Benali}, \citenamefont {Fert{\'e}}, \citenamefont {Paquier},
		\citenamefont {Pradines}, \citenamefont {Assaraf}, \citenamefont {Reinhardt},
		\citenamefont {Toulouse}, \citenamefont {Barbaresco}, \citenamefont {Renon},
		\citenamefont {David}, \citenamefont {Malrieu}, \citenamefont {V{\'e}ril},
		\citenamefont {Caffarel}, \citenamefont {Loos}, \citenamefont {Giner},\ and\
		\citenamefont {Scemama}}]{QP2}%
	\BibitemOpen
	\bibfield  {author} {\bibinfo {author} {\bibfnamefont {Y.}~\bibnamefont
			{Garniron}}, \bibinfo {author} {\bibfnamefont {K.}~\bibnamefont {Gasperich}},
		\bibinfo {author} {\bibfnamefont {T.}~\bibnamefont {Applencourt}}, \bibinfo
		{author} {\bibfnamefont {A.}~\bibnamefont {Benali}}, \bibinfo {author}
		{\bibfnamefont {A.}~\bibnamefont {Fert{\'e}}}, \bibinfo {author}
		{\bibfnamefont {J.}~\bibnamefont {Paquier}}, \bibinfo {author} {\bibfnamefont
			{B.}~\bibnamefont {Pradines}}, \bibinfo {author} {\bibfnamefont
			{R.}~\bibnamefont {Assaraf}}, \bibinfo {author} {\bibfnamefont
			{P.}~\bibnamefont {Reinhardt}}, \bibinfo {author} {\bibfnamefont
			{J.}~\bibnamefont {Toulouse}}, \bibinfo {author} {\bibfnamefont
			{P.}~\bibnamefont {Barbaresco}}, \bibinfo {author} {\bibfnamefont
			{N.}~\bibnamefont {Renon}}, \bibinfo {author} {\bibfnamefont
			{G.}~\bibnamefont {David}}, \bibinfo {author} {\bibfnamefont {J.~P.}\
			\bibnamefont {Malrieu}}, \bibinfo {author} {\bibfnamefont {M.}~\bibnamefont
			{V{\'e}ril}}, \bibinfo {author} {\bibfnamefont {M.}~\bibnamefont {Caffarel}},
		\bibinfo {author} {\bibfnamefont {P.~F.}\ \bibnamefont {Loos}}, \bibinfo
		{author} {\bibfnamefont {E.}~\bibnamefont {Giner}}, \ and\ \bibinfo {author}
		{\bibfnamefont {A.}~\bibnamefont {Scemama}},\ }\href {\doibase
		10.1021/acs.jctc.9b00176} {\bibfield  {journal} {\bibinfo  {journal} {J.
				Chem. Theory Comput.}\ }\textbf {\bibinfo {volume} {15}},\ \bibinfo {pages}
		{3591} (\bibinfo {year} {2019})}\BibitemShut {NoStop}%
	\bibitem [{\citenamefont {Chakravorty}\ \emph {et~al.}(1993)\citenamefont
		{Chakravorty}, \citenamefont {Gwaltney}, \citenamefont {Davidson},
		\citenamefont {Parpia},\ and\ \citenamefont
		{p~Fischer}}]{ChaGwaDavParFro-PRA-93}%
	\BibitemOpen
	\bibfield  {author} {\bibinfo {author} {\bibfnamefont {S.~J.}\ \bibnamefont
			{Chakravorty}}, \bibinfo {author} {\bibfnamefont {S.~R.}\ \bibnamefont
			{Gwaltney}}, \bibinfo {author} {\bibfnamefont {E.~R.}\ \bibnamefont
			{Davidson}}, \bibinfo {author} {\bibfnamefont {F.~A.}\ \bibnamefont
			{Parpia}}, \ and\ \bibinfo {author} {\bibfnamefont {C.~F.}\ \bibnamefont
			{p~Fischer}},\ }\href {\doibase 10.1103/PhysRevA.47.3649} {\bibfield
		{journal} {\bibinfo  {journal} {Phys. Rev. A}\ }\textbf {\bibinfo {volume}
			{47}},\ \bibinfo {pages} {3649} (\bibinfo {year} {1993})}\BibitemShut
	{NoStop}%
	\bibitem [{\citenamefont {Davidson}\ \emph {et~al.}(1991)\citenamefont
		{Davidson}, \citenamefont {Hagstrom}, \citenamefont {Chakravorty},
		\citenamefont {Umar},\ and\ \citenamefont
		{Fischer}}]{DavHagChaMeiFro-PRA-91}%
	\BibitemOpen
	\bibfield  {author} {\bibinfo {author} {\bibfnamefont {E.~R.}\ \bibnamefont
			{Davidson}}, \bibinfo {author} {\bibfnamefont {S.~A.}\ \bibnamefont
			{Hagstrom}}, \bibinfo {author} {\bibfnamefont {S.~J.}\ \bibnamefont
			{Chakravorty}}, \bibinfo {author} {\bibfnamefont {V.~M.}\ \bibnamefont
			{Umar}}, \ and\ \bibinfo {author} {\bibfnamefont {C.~F.}\ \bibnamefont
			{Fischer}},\ }\href@noop {} {\bibfield  {journal} {\bibinfo  {journal} {Phys.
				Rev. A}\ }\textbf {\bibinfo {volume} {{44}}},\ \bibinfo {pages} {7071}
		(\bibinfo {year} {1991})}\BibitemShut {NoStop}%
	\bibitem [{\citenamefont {Schraivogel}\ \emph {et~al.}(2021)\citenamefont
		{Schraivogel}, \citenamefont {Cohen}, \citenamefont {Alavi},\ and\
		\citenamefont {Kats}}]{SchCohAla-JCP-21}%
	\BibitemOpen
	\bibfield  {author} {\bibinfo {author} {\bibfnamefont {T.}~\bibnamefont
			{Schraivogel}}, \bibinfo {author} {\bibfnamefont {A.~J.}\ \bibnamefont
			{Cohen}}, \bibinfo {author} {\bibfnamefont {A.}~\bibnamefont {Alavi}}, \ and\
		\bibinfo {author} {\bibfnamefont {D.}~\bibnamefont {Kats}},\ }\href {\doibase
		10.1063/5.0072495} {\bibfield  {journal} {\bibinfo  {journal} {J. Chem.
				Phys.}\ }\textbf {\bibinfo {volume} {155}},\ \bibinfo {pages} {191101}
		(\bibinfo {year} {2021})}\BibitemShut {NoStop}%
\end{thebibliography}

%

 \end{document}